\documentclass[conference]{IEEEtran}
\usepackage{blindtext, graphicx}

\ifCLASSINFOpdf

\else

\fi

\begin{document}

\title{On Small-World Networks: Survey and Properties Analysis}

\author{\IEEEauthorblockN{Alaa Eddin Alchalabi}
\IEEEauthorblockA{\emph{Graduate School of Electronics and Computer Engineering}\\
Istanbul Sehir University, Istanbul, Turkey\\
alaaalchalabi@std.sehir.edu.tr}
}

\maketitle

\begin{abstract}
Complex networks has been a hot topic of research over the past several years over-crossing many disciplines, starting from mathematics and computer science and ending by the social and biological sciences. Random graphs were studied to observe the qualitative features they have in common in planetary-scale data sets which helps us to project the insights proven to real-world networks.

In this paper, We survey the particular case of small-world phenomena and decentralized search algorithms. We start by explaining the first empirical study for the “six degrees of separation” phenomenon in social networks; then we review some of the probabilistic network models based on this work, elaborating how these models tried to explain the phenomenon's properties, and lastly, we review few of the recent empirical studies empowered by these models. Finally, some future works are proposed in this area of research.\\

\end{abstract}


\IEEEpeerreviewmaketitle

\section{Introduction}
Recently, the study of complex networks has emerged in various range of disciplines and research areas. The World Wide Web has revolutionized the way we deal with everything one deals in daily life. Computer scientists were curious to find a way to handle the wheel of controlling the complexity and the enormous growth of the Internet. Social networks' data scale is unpredictably uncontrollable by social scientists. The biological interactions in cell's metabolism are expected to define its pathways and could provide insights to biologists [13]. The urge a new born science is needed in order to be able to manipulate networks before networks manipulate our needs [8]. \\

The study of complex networks evolved since the study of randomly generated graphs by Erdos and Renyi  [4], and the appearance of a large-scale network data had leashed tremendous work in multi-disciplinary areas including the real and the virtual world [13]. The efforts were put to describe the properties of random graphs in large networks which raised more and more technical questions to be answered. To mimic real-networks, a randomly produced stylized network model is adopted in order to generalize the resulting conclusions and  properties onto real-networks. Simple models fails to capture the complexity of a realistic network's structure and features offering a strong mathematical basis which futures investigations can be build upon.    \\

In the next sections of this paper, we survey the ``small-world phenomenon'' and few related  problems. We start with the famous psychologist Stanley Milgram's social experiment, that captures the main aspects of the phenomenon [11], [14]; we review few of the models based on random graphs that tries to explain the problem [7], [9], [12], [15], [16]; and then we  mention recent work that has applied the traditional insights of these models on large data sets extracted from famous web applications [2], [10]. Lastly, some suggested further extensions to small-world networks are discussed, along with some future works followed by their relevance to this field.

\section{small-world phenomenon}

The small-world phenomenon has recently been the hot topic of both theoretical and practical research, and it has been given huge attention from most, if not all, multi-disciplinary researchers. The term ``small-world'', linked by all means to the ``short chains of acquaintances'', or the ``six degrees or separation'' [5][6][16], refers to the human social network's graph; where nodes replaces people, and edges between two nodes mimic if the two corresponding persons know each other on a first-name basis [8]. The graph is described to be a ``small-world'' because of the fact that any two random pairs of nodes are separated by relatively a small number of nodes, generally less than 6. Although the first-name basis rule is a bit naive for an edge definition, the resulted graph behaves as a real-world network.  \\

Small-world networks are of great importance because of adoption to the limitations of either of the end extreme networks types; random networks and regular lattices. Small-world networks proved their ability to be used as frameworks for the study of interaction networks of complex systems [1].\\

The most important key of the small-world study is to prove the hypothesis that assumes the qualitatively shared structure among a variety of networks across variant fields. A common property arises in large networks which is the existence of short paths between most of the nodes pairs although nodes in network have a high degree of clustering.  Nodes can also be reached and navigated with no need of universal understanding of the whole network. Such properties contributed in describing large-scale social networks behavior, and additionally, they gave important insights to create applications to the internal design of decentralized peer-to-peer systems. \\

\subsection{Milgram's Experiment}
Stanley Milgram, the famous social psychologist, made an experiment to measure people connectivity in the USA in the 1960s and to test the small-world property [11][14]. The experiment questions the probability of an arbitrarily two selected individuals from a large data set to know each other in person. A target person was selected in the state of Massachusetts who was a Boston stockbroker, and 296 arbitrarily selected individuals as ``starting persons'' from Nebraska and Boston were asked to generate acquaintance chains to the stockbroker in Boston. The selected group were given a document of the described study with the target's name. The senders were asked to choose a recipient which they think that he/she will contribute to carry the message to the target person in the shortest way possible. The concept of ``roster'' was introduced to prevent the message goes back to a previous sender ``loop'' and to track the number of node the message reached. \\

The results of the experiment were quite astonishing. Sixty-four chains made their way to the target person. The mean number of the intermediaries was 5.2 with median of 6. Boston starting chains exhibits shorter range chains than Nebraska's chains. Additional experiments by Korte and Milgram proved that these numbers are quite stable [14].\\

Some comments on Milgram's experiments exhibit the inability of this model to be generalized to larger networks. Varying the information about the target person might affect the decisions taken by senders, and here we meant psychological and sociological factors take place.\\

\section{Small-world based empirical models}

\subsection{Watts and Strogatz's Model}
Watts and Strogatz came up with a model that aims to explain the small-world property. After Bollobas and de la Vega [3] introduced the theorem which proves the logarithmic property in the path length with respect to the number of nodes \emph{O(log n)} in small-world networks, Watts and Strogatz felt that there was something missing in the theorem. The model proposed considered small-world networks to be highly-structured with relatively a few number of random links added within. Long-range connections in this model plays a crucial rule in creating short paths through all nodes [15]. \\

The model adopts the idea of rewiring edges between nodes with a certain probability to generate a graph. The probability allows the change between regular networks (p=0) and random networks (p=1). The model starts by generating a ring of n connected nodes (average degree of k). Then, the rewiring of each edge with the probability p and the landing node is also chosen randomly. The clustering coefficient \emph{$(C_p)$}  is a measure which reflects the fraction of the connected neighbours to a node in a graph compared to all possible connections of the neighboured averaged over all nodes [15]. \\

The results derived for a regular graph (p = 0) a highly clustered (\emph{$C\sim$}3/4) and path length \emph{$L\sim n/2k$} where \emph{$n>k>ln(n)>1$} should be chosen. For random graphs (p = 1) the resulted network is poorly with a low clustering coefficient (\emph{$C\sim k/n$}) and a small path length \emph{$L\sim ln(n)/ln(k)$}. Their research also included three empirical real-world examples of small-world models, and their main finding was that the average path length of the chosen example was slightly higher than the random model, while the clustering coefficient had clearly a much higher value than the random model. Using their results, they reasoned how infectious disease spread rapidly in small-world societies.\\

Some drawbacks of Watts and Strgatz's model is that it cannot be generalized to all small-world models. Some extended works by other scientists tried to fill in gaps. \\

\subsection{Classes of Small-World Networks}

Due to the limited vision of the Watts and Strogatz model, new explanation was needed. Trying to look at the dilemma from another prospective, Amaral et al. tried to classify small-world networks to three classes reporting an empirical study of real-world networks [1]. The study covers mainly the statistical properties of real-world networks, and it was enough to prove the existence of three classes of small-world networks: Scale-free, broad-scale, and single scale[1].\\

\subsubsection{Scale-free}
The networks which is characterized by a vertex connectivity distribution which decays as a power  law.\\
\subsubsection{Broad-scale} The networks characterized by a connectivity distribution that has a power-low region and followed by a sharp cutt-off.\\
\subsubsection{Single-scale}The networks characterized by a connectivity distribution with a fast decaying tail.\\

The research also gave an answer to why such taxonomy exist, and they reasoned that by mentioning two types of factors. The first factor was the aging of the vertices, which in time old nodes will stop being as effective in the network, and an example of that can be the actors network. The second type of factor was the cost of adding new links to the vertices which limited by the vertex capacity. An example of this can be the airports map where placing too many vertices are pricey and not practical.\\

\subsection{Kleinbergs's Algorithmic Prospective}

Kleinberg's way of explaining small-world properties was a bit close to Watts and Strogatz but with slight differences [7]. Kleinberg used a \emph{n x n} grid of nodes to represent the network, and to add the small-world flavour, a number of long-range connection edges were added and not rewired. After adding the edges, the probability of connecting two random vertices (v,w) is proportional to \emph{$1/d(v,w)^{q}$} where q is the clustering coefficient [9].\\

Kleinberg came up with theorems to quantify the decentralized algorithms' delivery time which generalized the results in [3] by Bollobás and de la Vega of the logarithmic behavior of short pathways in networks. He proved that the time needed is not always logarithmic but it depends on other parameters. A new parameter was introduced ($\alpha$ \textgreater=0) that controls the long-range connections. Interestingly, the delivery time varies depends on $\alpha$ as follows:\\

\subsubsection{For 0 \textless $\alpha$ \textless 2} the delivery time of any decentralized algorithm in the grid-based model is $\Omega$ ($n^{(2-a)/3}$).

\subsubsection{For $\alpha$ = 2} the delivery time of any decentralized algorithm in the grid-based model is O($log^{2} n$).

\subsubsection{For  $\alpha$ \textgreater 2}  the delivery time of any decentralized algorithm in the grid-based
model is $\Omega$ ($n^{(a-2)/(a-1)}$) [7].\\

\section{Recent real-world Empirical experiments }

\subsection{Dodds, Muhammad, and Watts Experiment}

Dodds et al. tried to mimic Milgram's experiment with the electronic messaging systems. Around 60,000 randomly selected email users attempted to reach 18 target persons in 13 different countries.\\

The findings were quite unexpected. Successful social chains passed through intermediate to weak strength ties [12]. This finding proves that the highly connected hubs' effect is negligible. The attrition of message chains showed that messages could make it to the target in a median of five to seven. The cool fact about attrition rate was the constancy of its value for a certain period of time. The 384 completed chains (out of 24,163) had an average chain length of 4.05. This number was considered misleading by the authors, which made them evaluate the experiment using new metrics. \\

The general results showed that the the network structure alone is not enough to interpret the network. Actions and perceptions of the individuals are big contributors. \\

\subsection{Leskovec et al. on a Large Instant-Messaging Network }

Leskovec et al. presented a study in 2003 which captured a month of communication activities within the Microsoft Messenger instant-messaging system [10]. The data set contained about 30 million conversations between 240 million people, and a graph was constructed containing 180 million nodes and 1.3 billion undirected edges. The network represents accounts that were active during June 2006. \\

The resulted average path length among users was 6.6 with 6 being the median. The results showed that users with similar age, language, and location tend to communicate with each other. Users with different genders tend to communicate more and also for longer conversations [10]. Conversations tends to decrease with the increase in the distance. However, communication chains through relatively long distances tend to carry longer conversations [10]. \\

\subsection{Bakhshandeh's Degrees of Separation in Twitter }

Bakhshandeh et al. did an interesting analysis to identify the degree separation between two Twitter users.  They used a new search technique to provide near-optimal solutions more optimized than greedy approaches [2]. The average separation path length was 3.43 between any two random Twitter users which required 67 requests from Twitter, while the near-optimal was 3.88 using only 13.3 requests on average. Surprisingly, Twitter’s 3.43 degree of separation is small and the reason they have claimed was the indicative of changing social norms in the modern connected world. \\

\section{Further Extensions and future works}

There is no doubt that small-world networks are still and will still be a hot research topic  due to its nature. In this section, we would like to propose some ideas for future extensions which might propose solutions for unanswered or vaguely answered questions. \\

Introducing machine learning techniques to small-world networks, in my opinion, is a good idea. Constructing networks should be smart enough in order to be controllable not only interpretable. Small-world networks could be build to mimic the brain neural-map which might give us more insight on how the human brain works. ML techniques can be also used to conserve the ``six degrees of separation rule'' or even to break it which completely depends on the application.  \\

Introducing local reference nodes in such networks could be a new idea to be implemented. Reference nodes could have some regional knowledge about the surrounding nodes. They can control the ``hubs'' and determine how new links can be distributed among reference nodes. We can think of routers as examples. The uniqueness of the node is somehow unrealistic for some applications, and that shows the urge of introducing a new concept. \\

\section{Conclusion}
At this paper, we discussed the famous phenomenon of small-world networks and its importance in various areas. Few of the small-world driven models were surveyed. Then recent real-world experiments in the context of complex networks were mentioned. Later, further extensions and future works were proposed. In the future,  we will try to implement that the suggested ideas practically on a given data set. By taking into account their bros and cons, the ideas will be later evaluated against the other state-of-art implementations.  \\

\appendices

\ifCLASSOPTIONcaptionsoff
  \newpage
\fi

\begin{IEEEbiography}[{\includegraphics[width=1in,height=1.25in,clip,keepaspectratio]{picture}}]{John Doe}
\blindtext
\end{IEEEbiography}


\begin{thebibliography}{1}

\bibitem{1}
Amaral, Luıs A. Nunes, et al. "Classes of small-world networks." Proceedings of the national academy of sciences  97.21 (2000): \hskip 1em plus
  0.5em minus 0.4em\relax 11149-11152. 
  
\bibitem{2}
Bakhshandeh, Reza, et al. "Degrees of separation in social networks." Fourth Annual Symposium on Combinatorial Search. \hskip 1em plus
  0.5em minus 0.4em\relax 2011. 



\bibitem{3}
Bollobas, B., de la Vega,W. F., The diameter of random regular graphs. \hskip 1em plus
  0.5em minus 0.4em\relax
(1982), 125–134.

\bibitem{4}
Erdos, P., and Renyi, A., On the Evolution of Random Graphs. Magyar Tud. Akad. Mat. Kutató Int. Közl. \hskip 1em plus
  0.5em minus 0.4em\relax 5 (1960), 17–61.

\bibitem{5}
Fass, Craig, Mike Ginelli, and Brian Turtle. Six Degrees of Kevin Bacon. Plume Books, \hskip 1em plus
  0.5em minus 0.4em\relax 1996.

\bibitem{6}
Guare, John. Six degrees of separation: A play.\hskip 1em plus
  0.5em minus 0.4em\relax Vintage, 1990.

\bibitem{7}
J. Kleinberg. The small-world phenomenon: An algorithmic perspective. Proc. 32nd ACM Symposium on Theory of Computing, \hskip 1em plus
  0.5em minus 0.4em\relax 2000.

\bibitem{8}
J. Kleinberg. Complex Networks and Decentralized Search Algorithms. Proceedings of the International Congress of Mathematicians (ICM), \hskip 1em plus
  0.5em minus 0.4em\relax 2006. 

\bibitem{9}
Kleinberg, J. Navigation in a Small World. \hskip 1em plus
  0.5em minus 0.4em\relax Nature 406(2000), 845. 

\bibitem{10}
Leskovec, Jure, and Eric Horvitz. "Planetary-scale views on a large instant-messaging network." Proceedings of the 17th international conference on World Wide Web.\hskip 1em plus
  0.5em minus 0.4em\relax  ACM, 2008.

\bibitem{11}
Milgram, Stanley. "The small world problem." \hskip 1em plus
  0.5em minus 0.4em\relax Psychology today 2.1 (1967): 60-67.

\bibitem{12}
Peter Sheridan Dodds, Roby Muhamad, Duncan J. Watts. An Experimental Study of Search in Global Social Networks. \hskip 1em plus
  0.5em minus 0.4em\relax Science 301(2003), 827.

\bibitem{13}
Strogatz, Steven H. "Exploring complex networks." \hskip 1em plus
  0.5em minus 0.4em\relax Nature 410.6825 (2001): 268-276.

\bibitem{14}
Travers, Jeffrey, and Stanley Milgram. "An experimental study of the small world problem." \hskip 1em plus
  0.5em minus 0.4em\relax Sociometry (1969): 425-443.

\bibitem{15}
Watts, D. J. and S. H. Strogatz. Collective dynamics of 'small-world' networks. Nature 393:440-42(1998).

\bibitem{16}
Watts, Duncan J. Six degrees: The science of a connected age. \hskip 1em plus
  0.5em minus 0.4em\relax WW Norton and Company, 2004.


\end{thebibliography}
\end{document}